\documentclass[onecolumn,apjl]{emulateapj}

\input epsf
\usepackage{graphicx}
\newcommand{\beq}{\begin{equation}}
\newcommand{\beqa}{\begin{eqnarray}}
		  \newcommand{\eeq}{\end{equation}}
\newcommand{\eeqa}{\end{eqnarray}}

\newcommand{\lsim}{\lesssim}
\newcommand{\gsim}{\gtrsim}

\newcommand{\vect}[1]{\mbox{\boldmath${#1}$}} 
\newcommand{\lmk}{\left(}
\newcommand{\rmk}{\right)}
\newcommand{\lnk}{\left\{ }
\newcommand{\rnk}{\right\} }

\newcommand{\lla}{\left\langle}

\newcommand{\rra}{\right\rangle}

\newcommand{\vex}{{\vect x}}

\newcommand{\ven}{\vect n}

\newcommand{\vef}{{\vect f}}

\newcommand{\cA}{{\cal A}}

\begin{document}
\title{Non-Gaussianity test for discriminating gravitational wave backgrounds around 0.1-1Hz } 
\author{Naoki Seto}
\affil{National Astronomical Observatory, 2-21-1
Osawa, Mitaka, Tokyo, 181-8588, Japan
}

\begin{abstract}
We propose a non-Gaussianity test for gravitational wave backgrounds  by combining data streams of multiple detectors.  This simple method  allows us to check whether a detected background is "smooth" enough to be consistent with an inflation-type background, or is contaminated by individually undetectable weak burst signals. 
The proposed test would be quite useful for the Big Bang Observer or DECIGO whose primary target is a background from inflation at 0.1-1Hz where  gravitational wave bursts from supernovae of population III stars might become a troublesome foreground.

\end{abstract}
\keywords{gravitational waves---  early universe --- supernovae: general  }

\section{Introduction}
Since gravitational interaction is very weak, a gravitational wave (GW) background can serve as an invaluable fossil from early universe with almost no scattering and absorption during its propagation (Maggiore 2000). By analyzing the background, we might obtain crucial information to understand physics at extremely high  energy  scale, {\it e.g.} the inflation process.  However, overlap of multiple signals in data streams of detectors would become a basic aspect of GW astronomy, especially in the low frequency regime where number and variety of astrophysical sources would increase. This is partly because GW detectors have omni-directional sensitivity.  Therefore, for detecting a background from early universe, it is essential to disentangle diversified  signals contained in observed data.  For example, we need to detect and subtract individual astrophysical signals such as chirping binaries, from data streams (see {\it e.g.} Arnaud et al. 2007; Harms et al. 2008 for recent progress).

The primary target for the Big Bang Observer (BBO, Phinney et al. 2003) and  the DECihertz Interferometer Gravitational wave Observatory (DECIGO, Seto et al. 2001) is a background from inflation in the 0.1-1Hz band where a relatively deep window of GWs has been expected to be opened.
However, it was recently pointed out that burst-like GWs produced at supernova (SN) explosions of population-III (popIII) stars  might become a  problem for detecting an inflation background around 0.1-1Hz (Buonanno et al. 2004; Sandick et al. 2006; Suwa et al. 2007).  While a burst signal with a large amplitude might be handled in data analysis, a potential background composed by weak undetectable bursts would be a formidable obstacle for adequately identifying an inflation background. Furthermore, in contrast to regular waveforms ({\it e.g.} from binaries), it is often difficult to accurately model waveforms for burst signals and/or to predict their characteristic amplitudes beforehand. To deal with such a situation, we certainly need to study GW backgrounds with efficient quantification methods beyond the traditional simple measure $\Omega_{GW}$, the energy  density  of backgrounds (see also Finn et al. 1999; Drasco \& Flanagan 2002). In this paper we  propose a non-Gaussianity test for GW backgrounds in 0.1-1Hz band, and discuss its prospects with BBO, setting GWs from the popIII SNe as our fiducial burst model.

This paper is organized as follows: in \S 2 we briefly describe data streams of BBO with summaries for notations. The basic idea behind our non-Gaussianity  test is presented in \S 3.  In \S 4 we make a statistical evaluation for our method.

\section{Data Streams of BBO}
For BBO, two LISA-type units (armlength $L=5\times 10^4$km) would be used to form a David-star-like configuration. From each unit we can obtain three data streams $(A,E,T)$ and $(A',E',T')$ (Prince et al. 2002;  Corbin 
\& Cornish 2005; Seto 2006).   Among these six data,  $T$ and $T'$ modes are less sensitive to GWs, and we  neglect  them hereafter.  The data streams $A,E,A'$ and $E'$  can be effectively regarded as responses of simple L-shaped detectors around their optimal frequency $f_{opt}\sim 0.3$Hz with bandwidth $\delta f\sim f_{opt}$ (Seto 2006). In this paper we only deal with quantities made from the following two pairs; $A-A'$ or $E-A'$.  The orientation of the arms of the former pair is  aligned on the common detector plane of the two units,  but the latter is misaligned   by $45^\circ$ on the plane.

We model the data streams $s_X(t)$  ($X=A,E$ and $A'$) in terms of GW signal $H_X$ and detector noises $n_X$ as
$
s_X(t)=H_X(t)+n_X(t).
$
For analyzing the noises $n_X$, it is advantageous to work in the Fourier space. To clarify  our main points, we discuss signal analysis in the optimal band with neglecting details of frequency dependence  ({\it e.g.} replacing an integral $\int (\cdots) df $ with a product $(\cdots)_{f_{opt}}\times \delta f$).  In practice, this situation is approximately realized by applying a band-pass filter.
In order to take Fourier transformation, we decompose the data streams (total duration $T_{obs}$) into short segments of a given period $T_{seg} (\gsim f^{-1}_{opt})$, and assign  a label $M(=1,\cdots,T_{obs}/T_{seg})$  for each segment with chronological order.  Then we calculate
\beq
s_{XM}(f)=\int_{(M-1)T_{seg}}^{MT_{seg}} s_X(t)e^{2\pi i f t}dt=H_{XM}(f)+n_{XM}(f).\label{fourier}
\eeq
The number of relevant Fourier modes in a segment is  $\sim  T_{seg}\delta f \sim T_{seg} f_{opt}$.  In the Fourier transformations above, we implicitly assumed to apply an  adequate time window function to suppress leakage of underlying frequency components to nearby modes due to  finiteness of $T_{seg}$.

Hereafter, we only used the Fourier transformed quantities,  assuming that the detector noises $n_X$  are stationary, Gaussian, and have identical spectrum $S_N(f)$.  The assumption of Gaussian noises is not crucial for our method.
For the relevant pairs $(X,Y)=(A,A')$ or $(E,A')$, we also assume that their noises are independent (Phinney et al. 2003) with 
\beq
\lla n_{XM}(f) n_{YL}(f')^* \rra\sim \frac12 \delta_{ML}\delta_{XY} \delta_{ff'} T_{seg} S_{N}(f), \label{spectral}
\eeq
where the notation $\lla \cdots \rra$ represents to take an ensemble average.

Next we discuss detectors' responses to incoming GWs.
 We expand the metric perturbation due to a GW background at a time $t$ and position $\vex$ by 
\beq
h_{ij}(t,\vex)=\sum_{P=+,\times}\sum_{\vef} e^{2\pi i (\vef\vex+tf)} h_{In}(P,\vef) e_{ij}^P+{\rm c.c.}
\eeq
 (c.c.: complex conjugate) with three-dimensional wave vectors $\vef$, polarization bases $e_{ij}^P$  and a definition  $f\equiv |\vef|$.    
Response of a detector $X$ to an incident GW is characterized by the beam pattern function $F_X(\vef,P)$. In this paper, unless otherwise stated, we  study simple L-shaped interferometers with the long-wave approximation.  The explicit
form of the function $F_X$ is presented in the literature (Flanagan 1994; Allen \& Romano 1997).  For the background above, the signal $H_{XM}(\vef)$ in eq.(\ref{fourier}) is expressed as
\beq
H_{XM}(f)\simeq \sum_{P=+,\times}  \sum_{\vef\in B_f}\exp[{2\pi i (\vef\cdot\vex_X+ft_M)}]h_{In}(\vef,P)T_{seg} F_X(\vef,P)\label{inc}
\eeq
with a  shell-like three-dimensional frequency region $B_f$ corresponding to the observed frequency $f$.
 Then we have 
\beq
\lla H_{XM}(f) H_{YM}(f)^* \rra=\frac{8\pi}5 S_{GW,I}(f) \gamma_{XY} T_{seg},\label{cov}
\eeq
where the spectrum  ${ S}_{GW}(f)$ for the background has dimension  $[{\rm Hz^{-1}}]$ as for the detector noise spectrum $S_N(f)$, and is written with the normalized energy density $\Omega_{GW}$ by ${ S}_{GW}(f)=\frac{3H_0^2 }{32 \pi^3 G} f^{-3} \Omega_{GW}(f)$ ($H_0\simeq 70$km/sec/Mpc: the Hubble parameter). 
The overlap function $\gamma_{XY}(f)$  characterizes  magnitude of common responses  of two detectors $(X,Y)$ to isotropic and unpolarized backgrounds (Flanagan 1994; Allen \& Romano 1997). It is defined by 
\beq
\gamma_{XY}(f)\equiv \frac{5}{8\pi} \int_{Sphere} d\ven (F_X^+ F_Y^{+*}+ F_X^\times F_Y^{\times*}) e^{2\pi i  f \ven \cdot  (\vex_a-\vex_b)}.
\eeq
We have $\gamma_{EA'}=0$ due to geometrical symmetry and also $\gamma_{AA'}\sim1$ around the optimal band $f\sim 0.3$Hz of BBO (Seto 2006).  

\section{Non-Gaussianity Test}

If the signals $H_{XM}(f)$ in eq.(\ref{inc}) are made from superpositions of vast number of incoherent waves, they can be regarded as Gaussian variables from the central limit theorem. As a result,  they are characterized only by second moments, and  we have 
\beq
\lla (H_{XM}(f)  H_{YM}(f)^*)^2 
\rra=2\lla H_{XM}(f) H_{YM}(f)^* \rra^2  \propto \gamma_{XY}^2 \label{seki}
\eeq
 from the properties such as $\lla H_{XM}(f) H_{XM}(f)\rra=0$.

For a "coarse background" made by a relatively small number of freedom ({\it e.g.} popcorn noise due to supernova bursts),  the responses of detectors would be deviate from Gaussian.
Therefore, using the BBO pair $E$-$A'$  with the overlap function $\gamma_{EA'}=0$, we can check granularity of an isotropic background through the quantity $\lla (H_{EM}(f)  H_{A'M}(f)^*)^2
\rra$ that should vanish for a  Gaussian background {\it e.g.} that generated  at inflation. This is a key point in this paper.  For detectors with a finite overlap $\gamma_{XY}\ne 0$, we can  generalize this method  by introducing  combinations such as $\lla (H_{XM}(f)  H_{YM}(f)^*)^2
\rra-2 \lla H_{XM}(f)  H_{YM}(f)^*
\rra^2$ to subtract  Gaussian component (similar to the definition of the  Kurtosis parameter $\kappa_4$ for standard characterization of  non-Gaussianity (see {\it e.g.} Racine \& Cutler 2007)).  We do not pursue this direction further.  But the underlying approach proposed in this paper would be applicable to a network of ground based detectors.  For general detector configurations such as  the LIGO-VIRGO pair, we need the  subtraction scheme described above.

For statistical analysis with BBO,  we introduce the following two quantities made from the two pairs $A-A'$ and $E-A'$ respectively:
\beq
C_2\equiv \sum_M \sum_{f\in \delta f} s_{AM} (f) s_{A'M}(f)^*,~~~C_4\equiv \sum_M \sum_{f\in \delta f} (s_{EM} (f) s_{{A'}M}(f)^*)^2
\eeq
Here, the second summations  $\sum_{f\in \delta f}$ are for the Fourier modes (total number: $T_{seg}\delta f $) within a segment, and the first ones  $\sum_M$ are  for the segment label $M=(1,\cdots,T_{obs}/T_{seg}$).  The combination $C_2$ is used for traditional correlation analysis to measure $\Omega_{GW}$, while $C_4$ is our new probe for  non-Gaussianity of a GW background.  We evaluate their signal-to-noise ratios in the next section.

In this paper we have set   GWs from popIII SNe as our fiducial burst model.  Here we comment on other models.
Recently, several cosmological scenarios were proposed to produce intrinsically non-Gaussian GWs background in  early universe.  With a typical cosmological mechanism ({\it e.g.}  through preheating phases), GW background is generated by causal processes, when wavelength were comparable to or smaller than horizon size in order of magnitude sense (see {\it e.g.} Easther 
\& Lim 2006).  Therefore, even if the generated GWs have a  correlation structure, we  have a large number ($\gsim(fH_0^{-1})^2\sim 10^{36}(f/1{\rm Hz})^2$) of independent emission regions for GWs currently observed at frequency $f$,
and it would be difficult to directly probe the intrinsic non-Gaussianity for these typical models with BBO, due to the central limit theorem. But a background made by sparse cosmological events, such as GW bursts from cusps of cosmic strings might be an interesting target (Damour 
\& Vilenkin 2005).

\section{Burst Background}
\subsection{Derivation of Formulas}
In this section  we analyze a GW background made by a superposition of burst events from  single-species sources  that have an event rate $R$ and a  characteristic duration time $T_d$ for GW signal in the optimal  band.  We start our discussion in  a somewhat general way, and derive useful expressions for $C_2$ and $C_4$.     Our fiducial model (GWs popIII SNe)  will be examined later.

We assume a smooth spectral profile $\cA(f)^2$  $(\cA>0)$ for the GW emission throughout a burst event, and do not deal with regular waveforms with sharp frequency structures ({\it e.g.} monochromatic waves or chirping waves).   For  simplicity,   the emitted wave pattern observed at the Earth is assumed to have an  axisymmetric  profile described by  
\beq
(h_+,h_\times)\propto (\alpha_+(I),\alpha_\times (I)) 
\eeq
 ($I$: inclination angle, $\alpha_{+,\times}$: complex number)  in the coordinate system  that is apparently symmetric to the geometry of the source.  We also impose a  normalization condition;   $\frac12\int_0^\pi (|\alpha_+|^2+|\alpha_\times|^2) \sin I dI=1$.

 For our probes $C_2$ and $C_4$, we need to evaluate the second- and forth-oder moments  of the responses $H_{XM}$ of two  detectors induced by a burst event.  Each response $H_{XM}$ depends on the geometry  of the source relative to the detector. The geometry is characterized by four angular parameters;  the direction $(\theta,\phi)$ of the source on the sky and the orientation $(I,\psi)$ of its axis, and we have 
\beq
H_{XM}\propto \cA (F_X^+(\theta,\phi,\psi)\alpha_+(I) +F_X^\times(\theta,\phi,\psi)\alpha_\times(I)).
\eeq
Since these four geometric parameters are randomly distributed for extra-Galactic events, we define an averaging operator $[\cdots]_{an}$ with respect the direction and orientation of sources, and obtain  
\beq
[H_{XM}  H_{YM}^*]_{an}= \frac{\cA^2}5\gamma_{XY}. \label{2nd}
\eeq
 We define  the ratio
$Q\equiv [(H_{AM} H_{{ A'}M}^{*})^2]_{an}/([H_{AM}  H_{A'M}^*]_{an})^2$ for quantitative evaluation of the probe $C_4$. Its numerator is explicitly given as
\beq
[(H_{AM} H_{{A'}M}^{*})^2]_{an}=\frac{\cA^4}{630} \int_0^\pi \lmk 9\lnk |a_+|^4 +|a_\times|^4\rnk -34 | a_+ a_\times|^2+52 {\rm Re}[(a_+^* a_\times)^2] \rmk \sin I dI. \label{angle}
\eeq
Unless a weird cancellation occurs, the ratio $Q$ becomes order of unity.
At the end of this subsection, we will  explicitly demonstrate this  for our fiducial popIII SNe model.

With a relation $\gamma_{XX}=1$ for a self correlation, eq.(\ref{2nd}) shows that the angular average of the response function is $1/\sqrt{5}$. Therefore, 
   the  characteristic signal-to-noise ratio $SNR_{Bst}$   for individual burst with  a single detector  is evaluated as (see {\it e.g.} Segalis \& Ori 2001; Sago et al. 2004)
\beq
SNR_{Bst}\sim \frac{2 \lla  \cA^2\rra^{1/2}  (\delta f)^{1/2} }{5^{1/2} S_N^{1/2}}. \label{bst}
\eeq

Next we analyze the GW background formed by incoherent superposition of the bursts analyzed above, assuming that the typical signal-to-noise ratio $SNR_{Bst}$ is not larger than $O(1)$.  As the expected number of events in a segment is $RT_{seg}$, we get the background spectrum 
\beq
S_{GW,Bst}=\frac{R\lla \cA^2\rra}{8\pi}
\eeq
from expression (\ref{cov}) and the corresponding mean magnitude $\lla H_{XM} H_{YM}^*\rra=R T_{seg}\times \lla \cA^2\rra \gamma_{XY}/5$.  
Now we evaluate the signal-to-noise ratio $SNR_{Bst,C2}$ for the quantity $C_2$ under the condition  $T_{seg}\gg T_{d}$. 
If a segment $M$ contains a burst event, its averaged contribution to  $C_2$ is $\lla \cA^2\rra\gamma_{XY}T_{seg} \delta f /5$. After incoherent superposition of   $T_{obs} R (\gg 1) $ events during  observation time $T_{obs}$, we obtain  the expectation value (signal strength) $\lla C_2\rra= \lla \cA^2\rra\gamma_{XY} T_{seg} \delta f/5 \times (T_{obs} R)$.
Meanwhile, assuming independence of detector noises and the condition $S_N(f_{opt})>S_{GW}(f_{opt})$, the rms fluctuations for the product $s_{AM} s_{A'M}^*$ are  given by the detector noise spectrum as
$2^{-1/2} 2^{-1} S_N T_{seg}$.  Here the additional factor $2^{-1/2}$ is associated with the  projection operation of data to the expected phase direction (usually onto the real axis)  in the complex plane (Seto 2006). Taking into account
the total number of Fourier modes $(T_{obs}/T_{seg})\times (T_{seg} \delta f)=T_{obs} \delta f$, we obtain the rms fluctuations for $C_2$ as $2^{-3/2} S_N T_{seg} (T_{obs} \delta f)^{1/2}$,  and  its signal-to-noise ratio is given by 
\beq
SNR_{Bst,C2}=\frac{2}5  \frac{R\lla \cA^2\rra}{S_N} (2 T_{obs} \delta f)^{1/2}.\label{snr0}
\eeq
We can derive the same result for  $T_{seg}\lsim T_{d}$.
After replacing the product $\times \delta f$ with a frequency integral $\int df$ and using the spectrum $R \lla \cA^2\rra=8\pi S_{GW,Bst}$, the square value of this expression exactly matches with the  standard formula for correlation analysis given in the literature (Flanagan 1994; Allen \& Romano 1997).

In the same manner, we can derive $\lla C_4\rra=\lla \cA^4\rra Q R T_{seg} T_{obs}\delta f/25$ and obtain its signal-to-noise ratio as
\beq
SNR_{Bst,C4}=\frac{2^{3/2} \lla \cA^4\rra |Q| R (T_{obs} \delta f)^{1/2}}{25 T_{seg} S_N^2}  ~~~({\rm for} ~T_{seg}\gg T_d ) \label{snr1}
\eeq
with an explicit dependence on $T_{seg}$ in contrast to $C_2$.  Note that, here, independence of detector noises is an important requirement, but their Gaussianity is not essential.

For $T_{seg}\lsim T_d$, we  get the expectation value  $\lla C_4\rra=\lla \cA^4\rra Q R T_{seg}^2 T_{obs} T_d^{-1}\delta f/25$ with its  signal-to-noise ratio 
\beq
SNR_{Bst,C4}=\frac{2^{3/2} \lla \cA^4 \rra|Q| R (T_{obs} \delta f)^{1/2}}{25 T_{d} S_N^2} ~~~({\rm for} ~T_{seg}\lsim T_d ).\label{snr2}
\eeq

Eqs.(\ref{snr1}) and (\ref{snr2}) are re-expressed as
$SNR_{Bst,C4}=\frac{U}{4  \delta f \max\lnk T_{d}, T_{seg}\rnk} SNR_{Bst,c2} \cdot SNR_{Bst}^2$ with $U\equiv |Q| \lla \cA^4\rra/\lla \cA^2\rra^2=O(1)$.
With increasing the segment length $T_{seg}$ from its minimum $\sim f^{-1}_{opt}$,  we have a transition  from $\lla C_4\rra\propto T_{seg}^2$ to $\lla C_4\rra \propto T_{seg}^1$ at the point $T_{seg}\sim T_d$ where the signal-to-noise ratio $SNR_{Bst,C4}$ also starts to decrease from a constant given in eq.(\ref{snr2}) due to dilution of power. Therefore, if the signal $ C_4$ is detectable at $T_{seg}\sim f^{-1}_{opt}$,  we can estimate the duration time $T_{d}$ by  identifying the transition. In the following, we set $T_{seg}\sim f^{-1}_{opt}$ that will provide us with the maximum value of $SNR_{Bst,C4}$ for a burst background.

Now we focus our discussion on a burst model with $T_d\sim f^{-1}_{opt}$ ({\it e.g.}  for our fiducial popIII SNe model at 0.1-1Hz).  In this case, we have $SNR_{Bst,C4}=U \cdot SNR_{Bst,C2}\cdot SNR_{Bst}^2/4$, and this relation is very  suggestive.  Even if individual  burst  has signal-to-noise ratio $SNR_{Bst}$ less than detection threshold, there is an amplification factor $SNR_{Bst,C2}(\propto T_{obs}^{1/2})$ that increases with observational time $T_{obs}$ and enable us to statistically study the bursts.

With the parameters related to the bursts, we obtain $SNR_{Bst,C4}=(S_{GW,Bst})^2U/R\times 64\pi^2 (2T_{obs}\delta f)^{1/2} \delta f/(25S_N^2)$.  For a fixed background level $S_{GW,Bst}(\propto R\lla \cA^2\rra)$, the signal-to-noise ratio  $SNR_{Bst,C4}$ decreases for a higher event rate $R$ (corresponding to a smaller amplitude $\lla \cA^2\rra$).  This is  reasonable, considering that the background would become more Gaussian-like. If the bursts events are  supposed to be  the dominant sources of the total GW background around $f_{opt}$ and both $\lla C_2 \rra \propto R\lla \cA^2\rra$ and $\lla C_4 \rra \propto R\lla \cA^2\rra^2 U/T_d$ are measured, we can roughly estimate the event rate $R$ and the amplitude $\lla \cA^2\rra$ separately, assuming $U=O(1)$.  In addition to the estimated duration $T_d$, these will be basic information to disclose the nature of the burst sources.

As we commented earlier, the  ratio $Q\equiv [(H_{EM} H_{{ A'}M}^{*})^2]_{an}/([H_{AM}  H_{A'M}^*]_{an})^2$ becomes order of unity for typical burst waveforms.  Here we  demonstrate this  for our fiducial model: burst GWs from popIII SNe. In the band around 0.1-1Hz, the emitted waves are dominated by memory effects caused by anisotropic neutrino emissions at supernova explosions (Buonanno et al.  2004), and we put the GW waveforms as  $(\alpha_+,\alpha_\times ) \propto \lnk \sin^2 I (1+\frac35 q-\frac15q\sin^2I), 0\rnk$ (Epstein 1978).  The parameter $q(\lsim O(1))$ characterizes  asphericity of the  emission with  $q>0$ for a polar enhancement and $q<0$ for an equatorial enhancement. Compared with the standard expression,  we omitted an overall factor ($\propto q$) that can be absorbed into the power $\cA$.  For this model the ratio is given as $Q=0.99$ ($q=1$), $Q=1.02$ ($q=0$) and  $Q=1.10$ ($q=-1$).   We have limited our non-Gaussianity test only with the fixed pair $E-A'$. By combining results from geometrically different pairs, we can in principle, extract polarization information for burst signals.

\subsection{Prospects around 0.1-1Hz}
In this subsection we specifically discuss prospects of our non-Gaussianity test for popIII SNe GW background with BBO.
 Since the anisotropies of neutrino emissions from popIII SNe or the  formation rate of popIII stars are poorly understood,  the amplitude of the popIII background is currently quite uncertain.    With a parameter set for popIII SNe (redshift $z\sim15$,  emitted neutrino energy $E_\nu\sim 10^{55}$erg with mean isotropy $\lla q\rra\sim 0.03$ and event rate $R\sim 0.01{\rm sec^{-1}}$) extracted from Buonanno et al. (2004), we have $\Omega_{GW}\sim 4\times 10^{-16}$ at $f\sim 0.3$Hz.  Meanwhile, for  bursts  characterized by the background level $\Omega_{GW}$ and their rate $R$,  BBO has a sensitivity corresponding to 
\beq
SNR_{Bst}\sim 0.6\lmk \frac{\Omega_{GW}}{4\times 10^{-16}}\rmk^{1/2} \lmk \frac{R}{0.01{\rm sec^{-1}}}\rmk^{-1/2}
\eeq
 for individual bursts, and 
\beq
SNR_{Bst,C2}\sim 80 \lmk\frac{\Omega_{GW}}{4\times 10^{-16}}\rmk  \lmk \frac{T_{obs}}{10{\rm yr}}\rmk^{1/2},
\eeq
\beq
SNR_{Bst,C4}\sim 10 \lmk \frac{\Omega_{GW}}{4\times 10^{-16}}\rmk^{2} \lmk \frac{R}{0.01{\rm sec^{-1}}}\rmk^{-1}  \lmk \frac{T_{obs}}{10{\rm yr}}\rmk^{1/2}
\eeq
 for  the background with $T_{seg}\cdot \delta f \sim1$, $Q\sim1$ and $U \sim 1$.  Therefore, while identification of each burst might be difficult with small $SNR_{Bst}$, our method has potential to discriminate whether a background once detected is smooth enough and consistent with inflation origin.

An interesting question related to our non-Gaussianity test is whether we can separate smooth and burst contributions for the total energy  spectrum $\Omega_{GW}$.  To  estimate the latter component, we need the combination $R\lla \cA^2\rra$.  With our approach based on a forth-order moment, we can obtain a different combination $RQ\lla \cA^4\rra=(R \lla \cA^2\rra)^2\times U/R$.  If the burst rate $R$ is independently estimated {\it e.g.} with optical observation of popIII SNe, we can roughly estimate the burst component $\propto R\lla \cA^2\rra$ in the total spectrum $\Omega_{GW}$ by  introducing a model parameter $U=O(1)$.

The author would like to thank N. Kanda, S. Kawamura, M. Shibata, H. Tagoshi and T. Tanaka for simulating conversations. He also thanks the Aspen Center for Physics where part of this work was done, and acknowledges supports
from  Grants-in-Aid for Scientific Research of the Japanese Ministry
of Education, Culture, Sports, Science, and Technology
 20740151.



\end{document}